\documentclass[pra,twocolumn]{revtex4-2}
\usepackage{graphicx}
\usepackage{amsmath}
\usepackage{natbib}
\usepackage{epstopdf}
\usepackage{xcolor}

\newcommand{\n}{\nonumber}
\newcommand{\bn}{\begin{eqnarray}}
\newcommand{\en}{\end{eqnarray}}
\newcommand{\eml}{\end{multline}}
\newcommand{\bml}{\begin{multline}}

\begin{document}
\title{Benchmark for Quantum Teleportation with Non-Uniform Prior Distributions}
 \author{Tom\'a\v{s} Opatrn\'y,$^1$ Allison Brattley,$^{2}$ and Kunal K. Das$^{3,4}$}
  \affiliation{$^1$Department of Optics, Palacký University, 771 46 Olomouc, Czech Republic}
  \affiliation{$^2$Department of Physics, Yale University, New Haven, CT 06520, USA}
  \affiliation{$^3$Department of Physics and Astronomy, Stony Brook University, New York 11794-3800, USA}
  \affiliation{$^4$Department of Physical Sciences, Kutztown University of Pennsylvania, Kutztown, Pennsylvania 19530, USA}
\date{\today }

\begin{abstract}
Quantum teleportation should surpass maximum fidelity thresholds possible with local measurements and classical communications. Benchmarks have been established when states are drawn from a uniform distribution of qubits or  coherent states of harmonic oscillators. These are not applicable when there is prior knowledge about the input that skews the distribution. We determine the highest mean fidelity achievable for teleportation of states without using entanglement but drawn from a non-uniform prior distribution, specifically a von Mises - Fisher distribution. Using Bayesian methods, we derive benchmark functions for single qubit teleportation that utilize projective measurement on specific axes on the Bloch sphere. We demonstrate that similar results can be achieved with a coherent spin positive operator-valued measure (POVM), based upon heterodyne detection. These methods are then used to derive corresponding benchmarks for teleporting multi-qubit systems that we model as spin coherent states, covering the full range  of particle number $N\in [1, \infty)$. The results show that fidelity thresholds to declare quantum advantage can be significantly higher than is typically assumed with uniform distribution.
\end{abstract}

\maketitle

\section{Introduction}

Using pairs of separate entangled subsystems \cite{Bell} as a resource to remotely replicate a state is generally referred to as quantum teleportation.  The initial idea  \cite{Wootters1993} was promptly realized in experiments  \cite{bouwmeester_experimental_1997-1}.  Since then,  quantum teleportation has been demonstrated in a wide variety of systems \cite{hu_progress_2023}, where the fundamental qubits are photons \cite{marcikic_long-distance_2003, Polzik1998}, ions \cite{barrett_deterministic_2004,riebe_deterministic_2004}, and atomic states \cite{Jian-Wei_Pan-PNAS,Monroe_2009}.  There have been implementations in solid state systems \cite{pfaff_unconditional_2014,steffen_deterministic_2013}, with optomechanical systems \cite{fiaschi_optomechanical_2021}, and even with propagating microwaves \cite{fedorov_experimental_nodate}. Quantum teleportation of photons has been achieved over large distances, even between ground and satellite \cite{ren_ground--satellite_2017} and between distant nodes in a network \cite{hermans_qubit_2022}. There have been generalizations involving higher dimensional states \cite{Pan2019, PhysRevApplied.22.054045,lv_demonstration_2024}, quantum gates \cite{wan_quantum_2019,feng_chip--chip_2025}, and continuous variables \cite{Vaidman1994,BraunsteinKimble}.

In order to characterize a process as quantum teleportation, it is imperative to have some well-defined threshold that can be exceeded only if an entangled state is shared between the  separated locations involved. Otherwise, the process could be just a cleverly designed classical mechanism. Well-known examples of such benchmarks include the limit of $2/3$ for the mean fidelity of teleporting a random state of a qubit \cite{Bouwmeester01022000}, or the limit of $1/2$ for the mean fidelity of teleporting a randomly selected coherent state of a single-mode optical field \cite{Braunstein2000}. A limitation of such existing benchmarks is that the input states are drawn from completely flat distributions.  However, the states of real physical systems are very likely to follow distributions that are not uniform,  creating scenarios where there is  some prior knowledge about the state \cite{BOD-2026}. In such cases, the available benchmarks are no longer adequate, since by using only local measurements on the input state and classical communications, the receiving party could re-create the input states with a fidelity that can exceed these limits. A typical example is the fidelity of coherent states drawn from a Gaussian distribution corresponding to a thermal state with mean photon number $\langle n \rangle$: Using heterodyne detection on the input state and taking into account the prior knowledge, one can achieve  mean fidelity given by $\langle F \rangle = (\langle n \rangle + 1)/(2\langle n \rangle + 1)$ \cite{Braunstein2000,Hammerer05}. Whereas for $\langle n \rangle \gg 1$ this value approaches 1/2, for small  $\langle n \rangle$ the fidelity can be higher, even approaching 1 in the limit $\langle n \rangle \ll 1$.
There is therefore a necessity to generate reliable benchmarks for quantum teleportation of states of a qubit or of collective spins which are drawn from a nonuniform distribution, which defines the purpose of this paper.

While for harmonic oscillator coherent states with Gaussian prior distribution the results have been explored in previous literature \cite{Braunstein2000}, little is known for other systems.  Here, we consider the teleportation of a spin coherent state of $N$ qubits drawn from a prior distribution. We want to determine the maximum achievable mean fidelity for re-creating the input state at the receiving party provided no entanglement is shared. This can serve to benchmark the quantum advantage for such teleportation scenarios.  We find an analytical expression for the fidelity limit that defines the benchmark, in situations where the local measurements are generalizations of the standard heterodyne measurement for spin coherent states. We present evidence to support that this limit is very close to the mathematical limit based on projections on suitable bases. Our results can be used to assess and compare diverse quantum teleportation schemes that operate by sharing entangled collective spins, that spans the full range of system sizes from single qubit teleportation to the limiting case of an infinite number of particles.

The paper is organized as follows. In Sec.~\ref{SQubit}, we consider a single qubit that has a prior distribution peaked at a pole of the Bloch sphere, and use axis-specific projective measurements to derive entanglement-free fidelity thresholds that can substantially exceed accepted values that assume uniform distributions.  We develop a more practical positive-operator value measure (POVM) approach in Sec.~\ref{SecCollective} based on Bayesian statistics, and demonstrate that it yields maximum fidelities very close to the axis-specific benchmarks. We apply the POVM method to the more general case of teleporting multi-particle states in Sec.~\ref{SecConcl}, and present benchmarks that are applicable in the presence of non-uniform prior distributions. Section~\ref{SecConcl} summarizes our primary results and discusses outlooks for future work. The details of our analytical derivations are shown in two Appendices, \ref{AppOneQubit} and \ref{AppFidOpt}.

\section{Single qubit} \label{SQubit}

Assume a qubit prepared in a randomly chosen pure state
\begin{eqnarray}
|\theta,\phi\rangle = \cos \frac{\theta}{2}|0\rangle + e^{i\phi} \sin\frac{\theta}{2}|1\rangle,
\end{eqnarray}
and that the parameters $\theta, \phi$ follow a von Mises-Fisher (VMF) probability distribution, which is analogous to a thermal distribution but adapted to the compact space of a Bloch sphere,
\begin{eqnarray}
P(\theta, \phi) = \xi e^{\kappa \cos \theta},\quad \xi=\frac{\kappa}{4\pi \sinh \kappa},
\label{EProbtheta}
\end{eqnarray}
with $\kappa$ real and normalization constant $\xi$. For positive $\kappa$, the distribution is peaked around $\theta=0$, and for negative $\kappa$ around $\theta=\pi$.
The mean number of excitations is
\begin{eqnarray}
\langle n \rangle &=& \frac{1}{2}\left( 1-\coth \kappa + \frac{1}{\kappa} \right) .
\label{EbetaMeanN}
\end{eqnarray}
Our goal is to use a set of measurement outcomes to determine the values of $\theta, \phi$ of the prepared state.

\begin{figure}[t!]
\centering
\includegraphics[width=\columnwidth]{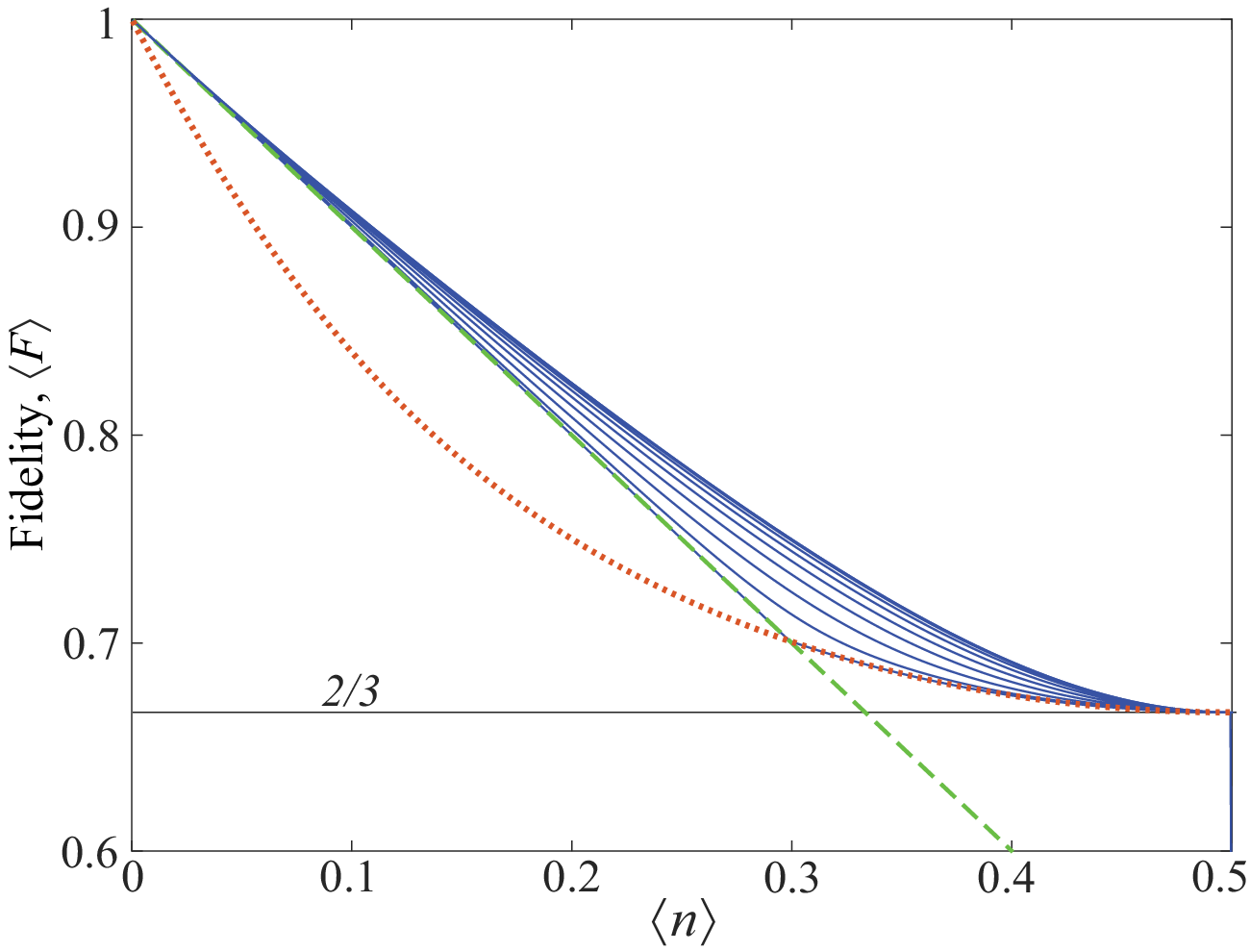}
\caption{Mean fidelity for no-entanglement teleportation of a single qubit using projection on different axes $\theta_0$ as a function of the mean number of excitations $\langle n \rangle$ of Eq. (\ref{EbetaMeanN}). The green dashed line corresponds to the ``do-nothing'' strategy as in Eq. (\ref{EFdonothing}). The red dotted line is teleportation using projection on the direction of prior knowledge, without averaging over the prior distribution. The top-most solid blue line corresponds to $\theta_0=\pi/2$, and the lines move down in steps of $\pi/16$. The bottom solid blue line overlapping with the green dashed and red dotted lines corresponds to $\theta_0=0$.}
\label{figAppen1}
\end{figure}

We set our baseline as the ``do-nothing'' strategy, where we simply guess the most probable state without doing any measurement. The mean fidelity is then
\begin{eqnarray}
\langle F \rangle_{\rm DN} &=& \frac{1}{2}\left( 1 + \coth \kappa - \frac{1}{\kappa} \right) = 1-\langle n \rangle.
\label{EFdonothing}
\end{eqnarray}
This is plotted as a dashed green line in Fig.~\ref{figAppen1} and Fig.~\ref{FigF1bit} as a reference for the optimized fidelities.

In the case of single qubit teleportation, we can do direct projective measurements, along an axis specified by Bloch sphere co-orodinates $(\theta_0,\phi_0)$. Due to the azimuthal symmetry of the VMF distribution, we can choose $\phi_0=0$ without loss of generality.  Such measurements will yield maximum information when the projection axis is orthogonal to the direction of the probability maximum at $\theta=0$ for $\kappa>0$ in Eq.~(\ref{EProbtheta}). Otherwise, there will be redundancy in the measured results, biased towards the most probable direction. Thus, knowing that the VMF distribution is peaked at the poles of the Bloch sphere, we will first assume a projection on $\sigma_x$ which is on the equator. If the measurement yields $\sigma_x = +1$, corresponding to having measured $\theta_M=\pi/2, \phi_M=0$, the probability of this result is
\begin{eqnarray}
P(x_+|\theta,\phi) &=& |\langle \theta_M, \phi_M|\theta,\phi \rangle |^2
=\frac{1}{2}(1+\sin \theta \cos \phi).
\end{eqnarray}
Using Bayesian inversion, we can find the conditional probability that the state was $|\theta, \phi\rangle$  under the condition that the measured result was $\sigma_x = +1$:
\begin{eqnarray}
P(\theta,\phi |x_+) &=& \xi e^{\kappa \cos \theta} (1+\sin \theta \cos \phi) .
\label{EqLikel}
\end{eqnarray}

Given this measurement $(\theta_M, \phi_M)=(\pi/2, 0)$, our task is to guess the parameters $\tilde{\theta},\tilde{\phi}$ that maximize the mean fidelity on averaging over all possible input values of  $\theta,\phi$. We consider guesses $(\tilde \theta,\tilde \phi =0) $, as the guessed values of $\tilde{\phi}$ should align with the measured value of $\sigma_x$. The fidelity of teleportation when guessing these angles for a given input state $|\theta, \phi\rangle$ is
\begin{eqnarray}
F(\theta,\phi) &=& |\langle \theta,\phi | \tilde \theta , \tilde \phi \rangle|^2 \nonumber \\
&=& \frac{1}{2}(1+\sin \tilde \theta \sin \theta \cos \phi + \cos \tilde \theta \cos \theta ).
\label{EqFid1}
\end{eqnarray}
The average fidelity over all input states given that a measurement yielded $\sigma_x = +1$ is then
\begin{eqnarray}
\bar F_+ &=& \int d\Omega\ F(\theta,\phi) P(\theta,\phi |x_+) \nonumber \\
&=& \frac{1}{2}\left[ 1 + \left( \coth \kappa - \frac{1}{\kappa} \right) \left( \frac{\sin \tilde \theta}{\kappa}+\cos \tilde \theta \right) \right] .
\label{EqFidelity}
\end{eqnarray}
\emph {Here, and elsewhere, we denote the angular integral in spherical co-ordinates by $\int d\Omega=\int_0^{2\pi}\int_0^{\pi}\sin \theta d\theta d\phi$.} The maximum mean fidelity is therefore achieved when
\begin{eqnarray}
\tilde \theta (+) = \arctan(\kappa^{-1}).
\label{Eq-Theta}
\end{eqnarray}
The same optimization can be done for the detection result  $\sigma_x = -1$, leading to $\tilde \phi = \pi$, $\tilde \theta (-) = \tilde \theta (+) = \arctan(\kappa^{-1})$,
and $\bar F_-=\bar F_+$. This follows from the azimuthal symmetry of the VMF distribution. Averaging over both possible detection results gives the optimized mean fidelity for a measurement axis $(\theta_0,\phi_0)=(\pi/2,\phi_0)$ on the equator,
\begin{eqnarray}
\langle F(\pi/2) \rangle &=&
P(x_+)\bar F_+ + P(x_-)\bar F_- \nonumber \\
&=&
\frac{1}{2}\left[1+ \left( \coth \kappa - \frac{1}{\kappa} \right) \sqrt{ 1 +\frac{1}{\kappa^2} } \right] .
\label{EqFMF}
\end{eqnarray}

For the general case where the measurement axis is arbitrary $(\theta_0,\phi_0)$ and not necessarily orthogonal to the direction of maximum probability of the prior distribution, a full derivation shown in Appendix A leads to the following expression for the optimized mean fidelity,
\begin{eqnarray}
\langle F(\theta_0)\rangle &=& \frac{1}{2} + \sqrt{A_+^2+B^2}+\sqrt{A_-^2+B^2},\n\\
A_{\pm} &=& \frac{1-2\langle n \rangle}{4}\left[ 1\pm \left( \frac{1}{1-2\langle n \rangle}-\frac{2}{\kappa} \right)\cos \theta_0 \right], \n\\
B &=& \frac{1-2\langle n \rangle}{4\kappa} \sin \theta_0.
\label{FidQubitThet0}\end{eqnarray}
In Fig.~\ref{figAppen1}, we plot the average fidelity $\langle F\rangle$ for different choices of projection axes, as computed with this expression. As one can check, this expression is maximized for $\theta_0 = \pi/2$ where it leads to Eq.~(\ref{EqFMF}).  This defines the upper bound in the figure, and marks the best teleportation fidelity for any $\langle n\rangle$ without using entanglement.

Projection along axes increasingly aligned with the axis of the prior distribution decreases the average fidelity, with the lower boundary corresponding to $\theta_0 = 0$, when Eq.~(\ref{FidQubitThet0})  reduces to $\langle F(0) \rangle = \frac{1}{2} + |A_+|+|A_-|$.  It is interesting to note that there is a cusp on that boundary, which occurs at $n_c = 1/(\kappa +2)$, where $A_-=0$; and using Eq. (\ref{EbetaMeanN}), we find that $n_c\simeq 0.3$.  For $n<n_c$, we find that lower boundary coincides with the dashed green line in Fig.~\ref{figAppen1} that marks the ``do-nothing'' fidelity given by Eq.~(\ref{EFdonothing}). But for $n>n_c$, the lower boundary coincides with the dotted red line in the figure, that we get if we \emph{neglect the prior distribution}, but still measure along its maximal axis and take the measured values as the guess $\tilde{\theta}, \tilde{\phi}$, with the associated mean fidelity given by
\bn
\langle F \rangle_{NP} = 1-\frac{1}{\kappa}(1-2\langle n \rangle). \label{F_NP}
\en
The two curves defined by Eqs.~(\ref{EFdonothing}) and (\ref{F_NP}) cross over precisely at $ n_c =1/(\kappa+2)$.  This means that if one projects the spins on the direction of prior knowledge, one receives no useful information for $\langle n \rangle \leq n_c$, as it is better to rely on the do-nothing strategy. Only for a more smooth prior distribution with $\langle n \rangle \geq n_c$ does taking into account the result of the measurement offer increases in mean fidelity. Regardless of the projection axis, the figure shows that in the limit $\kappa\to 0$ with $\langle n \rangle \to 1/2$, $\langle F\rangle\to 2/3$. In the opposite limit as $\kappa \to \infty$ with $\langle n \rangle \to 0$, which corresponds to sharply peaked distribution near $\theta =0$,  $\langle F\rangle \to 1$, as one would expect since we can guess with certainty the correct result $\theta =0$.

The main message of Fig.~\ref{figAppen1} is that in the presence of prior distribution, the often invoked 2/3 fidelity benchmark can be a significant underestimate for the maximum fidelity that can be achieved without using entanglement. This means that the fidelity threshold for claiming genuine quantum advantage in a teleportation scheme could be substantially higher, when the input state is picked from a distribution that is non-uniform.

\section{Coherent Spin POVM}
As we migrate away from a single qubit and consider teleportation of larger ensembles of particles, the projective measurements become impractical  for experiments as well as for computations \cite{BOD-2026, chaudhary_macroscopic_2025}, as one would have to project on basis states of a $2^N$-dimensional Hilbert space. Optimizing the basis as well as the guessed states inferred from the projection results would be a highly nontrivial task. Therefore, we explore here an alternative direction based on a positive operator-valued measure (POVM) detection scheme, which we first consider in the context of a single qubit before generalizing it to multi-particle states.

Let us consider a POVM with the set of operators $E(\theta_M,\phi_M) \equiv (2\pi)^{-1}|\theta_M,\phi_M \rangle \langle \theta_M, \phi_M|$ yielding the probability $p(\theta_M,\phi_M)={\rm Tr}\left[\rho E(\theta,\phi) \right]$, where $\rho$ is the relevant density matrix.
Thus, for a spin state oriented along $(\theta, \phi)$, the POVM scheme would detect a measured direction $(\theta_M, \phi_M)$ with probability $(2\pi)^{-1}|\langle \theta,\phi|\theta_M, \phi_M\rangle|^2$.
A particular physical realization for such a POVM has been proposed in \cite{DARIANO2002233}. Another option is to choose a random direction  $\theta_M$ and $\phi_M$ from a uniform distribution on a sphere, then project the state to be teleported on the coherent state centered at that angle, and determine whether it lies along this direction yielding $(\theta_M, \phi_M)$ or opposite yielding $(\theta_M + \pi, -\phi_M)$. In this section, we will show that even for a single qubit, the POVM scheme yields results that are near the upper bound for projective measurement on maximally separated axes; and we show this by deriving and comparing analytical expressions for the fidelities.

\begin{figure}[t]
\centering
\includegraphics[width=\columnwidth]{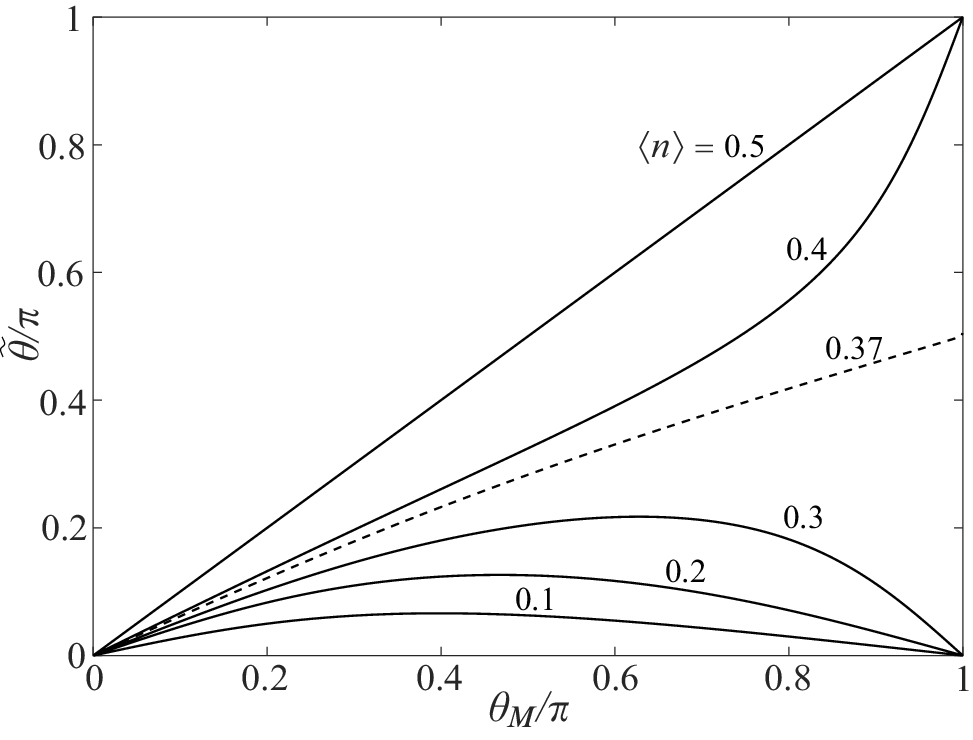}
\caption{Optimum estimated angle $\tilde \theta$ as a function of the detected angle $\theta_M$ for a single qubit POVM scheme, as in Eq. (\ref{EthetaOpt}), for various distributions of prior knowledge characterized by the values of $\langle n\rangle$. The dashed line corresponds to $n_0\simeq 0.37$, the turning point between monotonous increase and non-monotonous behavior.
}\label{FigOptTh}
\end{figure}

Assume the input state is $|\theta, \phi\rangle$, and measurement yields values $\theta_M$ and $\phi_M$. The conditional probability density for this result is
\begin{eqnarray}
&&P(\theta_M,\phi_M|\theta,\phi) = \frac{1}{2\pi}\cos^{2}\frac{\alpha}{2} \\
&=& \frac{1}{4\pi}\left[ 1 + \cos \theta_M \cos \theta + \sin \theta_M \sin \theta \cos (\phi_M -\phi)\right], \nonumber
\end{eqnarray}
where $\alpha$ is the angle between directions $(\theta, \phi)$ and $(\theta_M, \phi_M)$.
Bayesian inversion yields
\begin{eqnarray}
P(\theta,\phi|\theta_M,\phi_M) &=& \frac{P(\theta_M,\phi_M|\theta,\phi)P(\theta,\phi)}{P(\theta_M,\phi_M)},\n\\
P(\theta_M,\phi_M)&=& \int d\Omega\ P(\theta_M,\phi_M|\theta,\phi)P(\theta,\phi) \nonumber \\
&=& \frac{1+(1-2\langle n\rangle)\cos \theta_M}{4\pi}.
\end{eqnarray}

We now search for the optimum estimators $\tilde \theta = \tilde \theta(\theta_M,\phi_M,\kappa)$ and  $\tilde \phi = \tilde \phi(\theta_M,\phi_M,\kappa)$ that maximize the mean fidelity
\begin{eqnarray}
\bar F(\tilde \theta ,\tilde \phi) = \int d\Omega\ |\langle \tilde \theta, \tilde \phi|\theta, \phi  \rangle|^2 P(\theta,\phi|\theta_M,\phi_M).
\label{EFidBit}
\end{eqnarray}
Based on the axial symmetry of the VMF distribution, we know $\tilde \phi = \phi_M$ will maximize fidelity. For the optimum value $\tilde \theta$, we find
\begin{eqnarray}
\tan \tilde \theta = \frac{\sin \theta_M}{\kappa + \left( \frac{\kappa^3}{\kappa \cosh \kappa - \sinh \kappa}-2\right)\cos \theta_M} .
\label{EthetaOpt}
\end{eqnarray}
For $\theta_M=\pi/2$, this expression coincides with Eq.~(\ref{Eq-Theta}), and for $\kappa = 0$ it yields $\tilde \theta = \theta_M$. The dependence of $\tilde \theta$ on $\theta_M$ can be seen as a generalization of the concept of ``gain'' $g$, which multiplies the measured quadrature value to get the value of the displacement for continuous variable teleportation \cite{Cirac2005}. Here, the relationship goes beyond a simple proportionality. The dependence of $\tilde \theta$ on $\theta_M$ for various values of $\langle n \rangle$ is shown in Fig.~\ref{FigOptTh}. The denominator vanishing in Eq.~(\ref{EthetaOpt}) marks a critical point at $\kappa_0\simeq 0.81$, corresponding to $n_0\simeq 0.37$ using Eq.~(\ref{EbetaMeanN}).  For $\langle n\rangle \lesssim n_0$, the prior distribution is strongly peaked and the dependence becomes non-monotonous, having a turning point, as can be seen in Fig.~\ref{FigOptTh}.  In practical terms, this means that if $\theta_M$ is measured on the hemisphere opposite to that where the probability density is most localized, it would be better to guess $\tilde \theta$ aligned with that distribution to increase the chance of overlapping with the correct state.

\begin{figure}[t]
\centering
\includegraphics[width=\columnwidth]{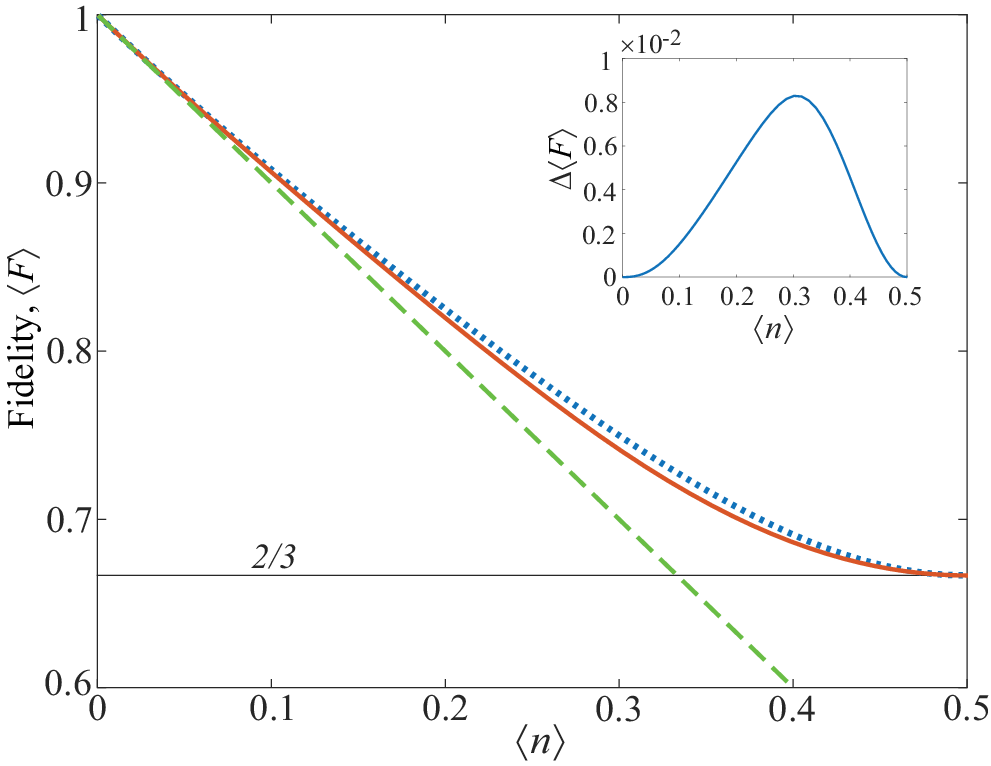}
\caption{Comparison of the mean fidelity for no-entanglement teleportation of a single qubit, with prior knowledge using POVM detection and projective measurement. The solid red line is the mean fidelity using a POVM as in Eq. (\ref{EqFOptPOVM}), and the blue dashed line uses projective measurement on $\sigma_x$ as in Eq. (\ref{EqFMF}). The inset displays the difference between the two lines. The ``do-nothing" strategy is included as a green dashed line for comparison. The mean fidelity limit of $2/3$ for no prior distribution is marked as a thin black line.
}\label{FigF1bit}
\end{figure}

Using the optimized estimation from Eq.~(\ref{EthetaOpt}) in the expression for fidelity in Eq.~(\ref{EFidBit}), we get
\begin{eqnarray}
\bar F(\theta_M) &=& \frac{1+\frac{1-2\langle n \rangle}{\kappa}\left\{ \kappa \cos \theta_M +
\sqrt{{\cal D}} \right\}}
{2\left[ 1 +(1-2\langle n \rangle)\cos \theta_M \right]},\\
 \text{with}\ {\cal D}&=& \left[ \kappa + \left( \frac{\kappa}{1-2\langle n\rangle}-2 \right)\cos \theta_M \right]^2 + \sin^2 \theta_M\n,
\end{eqnarray}
where $\langle n \rangle$ depends on $\kappa $ according to Eq.~(\ref{EbetaMeanN}).
Averaging this result over all possible detected values gives an analytical expression for the mean fidelity for the POVM scheme,
\begin{eqnarray}
\langle F \rangle &=& \int d\Omega_M\ \bar F(\theta_M) P(\theta_M,\phi_M)  \\
&=& \frac{1}{2}+ \frac{(1-2\langle n \rangle)^2}{16\kappa \sqrt{(3-6\langle n \rangle -\kappa)(1-2\langle n \rangle - \kappa)}} \nonumber \\
& \times & \left[2\left( {\cal B}_+{\cal C}_{+}  - {\cal B}_-{\cal C}_{-}  \right)
+  {\cal A} \ln \frac{\left({\cal C}_{+} +{\cal B}_+\right)\left({\cal C}_{-} -{\cal B}_-\right)}{\left({\cal C}_{+} -{\cal B}_+\right)\left({\cal C}_{-} +{\cal B}_-\right)} \right]\nonumber,
\label{EqFOptPOVM}
\end{eqnarray}
where
\begin{eqnarray}
{\cal A} &=& 1-\frac{\kappa^2 (1-2\langle n \rangle)^2}{(3-6\langle n \rangle -\kappa)(1-2\langle n \rangle - \kappa)} , \n\\
{\cal B}_{\pm} &=& \frac{\kappa(\kappa-2+4\langle n \rangle)}{\sqrt{(3-6\langle n \rangle -\kappa)(1-2\langle n \rangle - \kappa)}}
\n\\&&\pm \frac{\sqrt{(3-6\langle n \rangle -\kappa)(1-2\langle n \rangle - \kappa)}}{1-2\langle n \rangle}, \n\\
{\cal C}_{\pm} &=& \sqrt{{\cal A}+{\cal B}_\pm^2}.
\end{eqnarray}
We plot this mean fidelity as function of $\langle n\rangle$ in  Fig.~\ref{FigF1bit}, where we compare it with the best possible mean fidelity that can be achieved using projective measurements as given by Eq.~(\ref{EqFMF}). It shows the POVM fidelities closely follow those computed using projective measurements with axes, which, for single qubits, are maximally separated from the peak of the prior distribution. In the optimized projection scheme, we intentionally interrogate the orthogonal directions where the information on the input state is missing, whereas the POVM scheme also probes directions closer to the peak of the prior distribution, providing some redundant information. The resulting mean fidelity stems as an average of fidelities based on projections on random directions around the Bloch sphere. As the relative area near the equator is much bigger than the relative area near the poles, the contribution of near-perpendicular directions with higher mean fidelity is bigger than the contribution of near-parallel directions with lower fidelity. This means that the POVM scheme is a good estimate for the maximum mean teleportation fidelity achievable without utilizing quantum entanglement.

The single qubit case is actually the worst case for the coherent spin POVM, because with increasing number of qubits, $N$, we have $2^N$ optimal axes that are distributed across the sphere, instead of picking one orthogonal axis for a single qubit. So, as $N$ increases, a random choice of angle on the Bloch sphere is more likely to be close to one of the progressively more distributed optimal choices of axes to measure along; and in the limit $N\rightarrow \infty$, the two methods will converge. Figure~\ref{FigF1bit} shows that even for single qubit, the coherent spin POVM yields fidelities close to those obtained by projecting on an optimal measurement axis. With larger $N$, it will be closer still, and hence we will use coherent spin POVM measurements for teleporting systems with larger number or particles in the next section.


\section{Collective spin coherent state}
\label{SecCollective}

We now broaden our considerations to provide benchmarks that are applicable to multi-particle systems drawn from a non-uniform prior distribution. We will assume a spin coherent input state of $N$ qubits, and  aim to detect and re-create this state at a remote place without having an entangled state as a resource. For this purpose, it is useful to recall that this input state is equivalent to $N$ copies of a qubit prepared in the same state $|\theta,\phi\rangle$.

This can be put in the context of fidelity limits in previous work of Massar and Popescu  \cite{Massar-Popescu-PRL1995} where a uniform distribution was assumed, and the $N$-qubit state was utilized to make measurements to recreate a \emph{single} qubit. They derived a fidelity benchmark of such single-qubit remote replication to be $\langle F \rangle_{\rm 1-qubit} = \frac{N+1}{N+2}$ which, not surprisingly, approaches unity as $N\rightarrow \infty$. It can be easily shown that if instead the full $N-$particle state is re-created via $N$ measurements, an input state taken from a uniform distribution could be replicated with maximum fidelity $\langle F \rangle_{\rm N-qubit} = \frac{N+1}{2N+1}$, which falls off to the baseline value of $1/2$ as $N\rightarrow \infty$.

In contrast, we will assume the input state is picked from a non-uniform prior distribution, and our goal is to set a benchmark for replicating the complete $N$-particle state. We will use the coherent spin POVM discussed in the previous section,  which for the spin coherent input state $|\theta,\phi\rangle_N$ yields direction $\theta_M, \phi_M$ with a probability distribution proportional to the $Q$-function \cite{Mandel-Wolf},
\begin{eqnarray}
P(\theta_M,  \phi_M|\theta,\phi) = \frac{N+1}{4\pi}\cos^{2N}\frac{\alpha}{2} ,
\label{PthetMcondthet}
\end{eqnarray}
where $\alpha$ is the angle between directions  $\theta,\phi$ and $\theta_M, \phi_M$. As can be checked, if the probability distribution of directions $\theta,\phi$ is uniform, the mean fidelity of preparing the single qubit input state in the correct direction reduces to the expression in Ref.~\cite{Massar-Popescu-PRL1995}.

\begin{figure}[t!]
\centering
\includegraphics[width=\columnwidth]{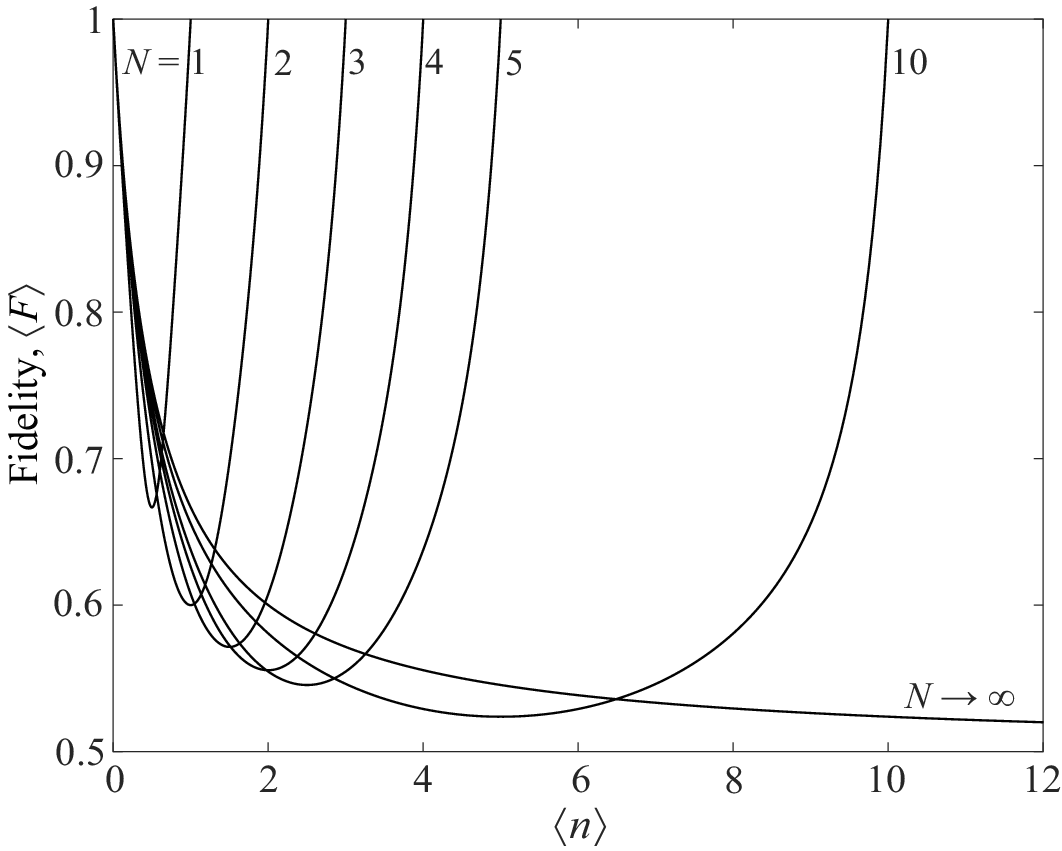}
\caption{Maximum mean fidelities achieved with no entanglement for different qubit numbers $N$ as a function of the mean number of excitations $\langle n\rangle$. The curves correspond to, from left to right, $N=1$, 2, 3, 4, 5, 10, and $N\to \infty$, as given by Eqs.~(\ref{EqBarFTildeThetaM}) and (\ref{EFlocal}), respectively.
}\label{FigFidelsBiased}
\end{figure}


Let us now assume that the spin coherent input states $|\theta,\phi\rangle_N$ are prepared with the VMF distribution $P(\theta,\phi)$ of Eq.~(\ref{EProbtheta}).
A particular spin coherent state $|\theta, \phi\rangle_N$ has a mean photon number $\bar n = N \sin^2 \frac{\theta}{2}$. Averaging over $P(\theta,\phi)$, we find
\begin{eqnarray}
\langle n \rangle &=& \frac{N}{2}\left( 1-\coth \kappa + \frac{1}{\kappa} \right) .
\end{eqnarray}
Denoting the measured values from the POVM scheme by $\theta_M$ and $\phi_M$, we first need to determine the optimal values
$\tilde \theta$ and $\tilde \phi$ such that the mean fidelity $\bar F (\theta_M,\phi_M)$ is maximized. This can be expressed as
\begin{eqnarray}
&&\bar F (\theta_M,\phi_M) = \\
&&\int d\Omega\  |\langle \tilde \theta(\theta_M,\phi_M,\kappa), \tilde \phi(\theta_M,\phi_M,\kappa)|\theta,\phi \rangle |^2
P(\theta,\phi|\theta_M,\phi_M),\n
\label{EMeanFidNprior}
\end{eqnarray}
where $P(\theta,\phi|\theta_M,\phi_M)$ is found using the Bayes rule as
\begin{eqnarray}
P(\theta,\phi|\theta_M,\phi_M) &=& \frac{P(\theta_M,\phi_M|\theta,\phi)P(\theta,\phi)}{P(\theta_M,\phi_M)}, \n\\
P(\theta_M,\phi_M) &=& \int d\Omega\ P(\theta_M,\phi_M|\theta,\phi)P(\theta,\phi).
\label{EPtMpM}
\end{eqnarray}
The functions $\tilde \theta(\theta_M,\phi_M,\kappa)$ and $\tilde \phi(\theta_M,\phi_M,\kappa)$ can be optimized analytically for maximum $\bar F (\theta_M,\phi_M)$, as discussed in detail in Appendix \ref{AppFidOpt}.
From the symmetry of the chosen prior distribution, we have $\tilde \phi = \phi_M$ and $\tilde \theta = \tilde \theta(\theta_M, \kappa)$.
The functions $\tilde \theta(\theta_M,  \kappa)$ for various $N$ are shown in Fig.~\ref{FigFThetaN}.

These functions once again generalize the concept of ``gain'' $g$ \cite{Cirac2005}: for large $N$ and $\theta_M \ll 1$, the estimated value is directly proportional to the measured value, $\tilde \theta \approx g \theta_M$ with $g$ being a function of $\langle n \rangle$ and $g=1$ for $\langle n \rangle = N/2$.
Averaging this over $P(\theta_M,\phi_M)$, we find the maximum achievable mean fidelity of the no-entanglement teleportation scheme. The full expression can be found in Appendix \ref{AppFidOpt} as Eq. (\ref{EqBarFTildeThetaM}). The resulting mean fidelities $\bar F_{N}(\langle n\rangle)$ are displayed in Fig.~\ref{FigFidelsBiased}, for particle numbers ranging from $N=1, \cdots 10$. They can be compared with the analytical expression \cite{Braunstein2000} when $N\to \infty$, also plotted in that same figure,
\begin{eqnarray}
\bar F_{\infty}(\langle n \rangle) = \frac{\langle n \rangle + 1}{2 \langle n \rangle + 1 }.
\label{EFlocal}
\end{eqnarray}
We see that for $\langle n\rangle < N/2$, our computed fidelities always lie below the corresponding  fidelities for the case of the idealized infinite--Hilbert space. This is significant in the context of specifying benchmarks for quantum teleportation, because for systems of finite $N$, our calculations show that quantum teleportation with the use of entanglement needs to beat a lower fidelity threshold than indicated by the infinite particle limit. But more significantly, Fig.~\ref{FigFidelsBiased} confirms our statement at the end of Sec.~\ref{SQubit} about prior distributions setting a higher threshold for quantum advantage, and broadens it to the teleportation of multi-particle states. As can be seen from the figure, $(N+1)/(2N+1)$ (=2/3 for a single qubit), which corresponds to the fidelity of teleporting a $N$ particle state chosen from a \emph{uniform} distribution, actually marks the \emph{minimum} of the optimized mean fidelities. This means that even for multiparticle states, having a prior distribution will generally raise the threshold fidelity that needs to be exceeded for a process to be labelled genuine quantum teleportation.

\section{Conclusion}\label{SecConcl}

We have shown here that teleporting a state taken from a non-uniform prior distribution will generally need to exceed a higher fidelity threshold to be considered genuine quantum teleportation, wherein entanglement provides a clear quantum advantage. We have determined new benchmarks to assess quantum teleportation when the input state is taken from a von Mises - Fisher prior distribution, a natural choice as the analog of a thermal distribution for compact spaces typical for qubits. Our results are applicable to the full range of particle number from single qubit to multiple particles that can tend to infinity. The methods we have developed here to set the benchmarks are based on general probabilistic considerations utilizing Bayesian statistics, and can be easily applied to other relevant prior distributions.

For a single qubit, the best strategy is to project the state on a direction perpendicular to the most probable direction of the prior distribution. Such projective measurements along a specific axis orientation however becomes impractical for states with larger numbers of particles. We have shown that by using a coherent spin POVM, one can achieve fidelities very close to those of the optimal projection strategy in the single qubit case. For spin coherent states of $N$ qubits, we derived analytical results for optimal POVM detections that reproduce the $Q$-function of the state. Determining the mathematical limit for fidelity achievable by local measurements and classical communication would require projections on a highly nontrivial basis of the $N$-particle states. Instead, we conjecture that the POVM fidelity limit that we obtain here will be close to the absolute maximum based on the observations that (i) these limits are very close even for $N=1$, which is actually the worst case for application of POVM, (ii) for $N\to \infty$ these limits coincide, as follows from the theorem of tight bound proven in \cite{Hammerer05}, and (iii) for arbitrary $N$ and no prior knowledge, the limits coincide with the results that assume uniform distribution \cite{Massar-Popescu-PRL1995}.

The benchmarks obtained here are close to the absolute maximum fidelity threshold for teleportation with no use of entanglement. Thus, if a scheme provides mean fidelities substantially higher than  Eq. (\ref{EqBarFTildeThetaM}), it is very likely that entanglement has been involved. On the other hand, if the scheme cannot generate fidelities above Eq. (\ref{EqBarFTildeThetaM}), the results could be reproduced by local measurements and classical communication. The POVM approach we utilize here to arrive at those benchmarks also has the benefit of being tractable for computation and practical for experiments. We derived our results here in the context of a specific class of teleported states: spin coherent states taken from a von Mises - Fisher prior distribution. Further research can apply our methods for benchmarking teleportation of a broader class of states, such as spin squeezed states, rotated Dicke states, or states with prior distributions with more complex structure. While the results for POVM detection schemes are expected to be very close to the ultimate fidelity bounds, finding them as well as defining optimal detection schemes will be interesting avenues to explore.

\begin{acknowledgments} This work was supported by the Czech Science
Foundation Grant No. 20-27994S for T. Opatrn\'y and by the NSF under Grant No. PHY-2011767 and PHY-2309025 for Kunal K. Das.  \end{acknowledgments}

\begin{widetext}
\appendix

\setcounter{figure}{0}
\renewcommand{\thefigure}{B\arabic{figure}}
\makeatletter
\def\fnum@figure{\figurename~\thefigure}
\makeatother

\section{One-qubit fidelity for projections on arbitrary direction}
\label{AppOneQubit}

Assume a qubit with a prior probability distribution given by Eq.~(\ref{EProbtheta}) and a projection scheme on an axis with direction $\theta_0$ and $\phi_0=0$. The projection gives a result ``+'' or ``$-$'', representing along or counter the chosen direction, respectively. The conditional probability for these results are
\begin{eqnarray}
P(\pm|\theta,\phi) = \frac{1}{2}\left(1\pm \sin\theta_0 \sin\theta \cos \phi \pm \cos \theta_0 \cos\theta  \right),
\end{eqnarray}
and Bayesian inversion yields
\begin{eqnarray}
P(\theta,\phi|\pm) & =& \frac{P(\pm|\theta,\phi)P(\theta,\phi)}{P(\pm)}, \\
{P(\pm)} &=& \int d\Omega P(\pm|\theta,\phi)P(\theta,\phi)  \nonumber \\
&=& \frac{1}{2}\left[1\pm  \left(\coth \kappa -\frac{1}{\kappa} \right) \cos \theta_0\right] = \frac{1}{2}\left[1\pm  \left(1-2\langle n\rangle \right)\cos \theta_0 \right] .
\end{eqnarray}
Depending on the projection result, let us choose the output state to be $|\tilde \theta_+,\tilde \phi_+\rangle$ or $|\tilde \theta_-,\tilde \phi_-\rangle$ with $\tilde \phi_+=0$, $\tilde \phi_-=\pi$, and $\tilde \theta_{\pm}$ chosen such as to maximize the mean fidelity
\begin{eqnarray}
\langle F \rangle = P(+)F_+  + P(-)F_-,
\end{eqnarray}
where
\begin{eqnarray}
F_{\pm} &=& \int d\Omega  |\langle \tilde \theta_{\pm},\tilde \phi_{\pm}|\theta,\phi\rangle|^2 P(\theta,\phi|\pm), \nonumber \\
|\langle \tilde \theta_{\pm},\tilde \phi_{\pm}|\theta,\phi\rangle|^2&=&  \frac{1}{2}\left(1+ \sin \tilde \theta_{\pm} \sin \theta \cos \phi + \cos \tilde \theta_{\pm} \cos\theta \right).
\end{eqnarray}
This yields
\begin{eqnarray}
P(\pm)F_{\pm} &=& \frac{\kappa}{16\pi \sinh \kappa} \int d\Omega  e^{\kappa \cos \theta}
\left(1\pm \sin\theta_0 \sin\theta \cos \phi \pm \cos \theta_0 \cos\theta  \right)
\left(1+ \sin \tilde \theta_{\pm} \sin \theta \cos\phi + \cos \tilde \theta_{\pm} \cos\theta \right) \nonumber \\
&=& \frac{1}{4}\pm \frac{1-2\langle n \rangle}{4}\cos \theta_0 + A_{\pm}\cos \theta_{\pm}\pm B\sin \theta_{\pm},
\end{eqnarray}
where
\begin{eqnarray}
A_{\pm} &=& \frac{1-2\langle n \rangle}{4}\left[ 1\pm \left( \frac{1}{1-2\langle n \rangle}-\frac{2}{\kappa} \right)\cos \theta_0 \right], \\
B &=& \frac{1-2\langle n \rangle}{4\kappa} \sin \theta_0.
\end{eqnarray}
The maximum values of $P(\pm)F_{\pm}$ are achieved for the choice of $\theta_{\pm}$ satisfying
\begin{eqnarray}
\tan \theta_{\pm} =  \frac{\pm B}{A_{\pm}},
\end{eqnarray}
which yields Eq.~(\ref{FidQubitThet0}) in Sec.~\ref{SQubit}.

\section{Fidelity optimization for $N$-qubit systems}
\label{AppFidOpt}

Since the prior probability $P(\theta,\phi) \equiv P(\theta)$ does not depend on $\phi$, we can use the azimuthal symmetry and take the best estimate of $\phi$ to be the measured value, i.e.,
\begin{eqnarray}
\tilde \phi(\theta_M,\phi_M,\kappa) = \phi_M
\end{eqnarray}
and
\begin{eqnarray}
P(\theta_M,\phi_M) &=& P(\theta_M,0) \nonumber \\
&=&  \int d\Omega P(\theta_M,0|\theta,\phi)P(\theta).
\end{eqnarray}
The task then is to find the function $\tilde \theta(\theta_M)$ such that the mean fidelity as given in Eq.~(\ref{EMeanFidNprior}) is maximized.
Since
\begin{eqnarray}
P(\theta_M,\phi_M|\theta, \phi )&=& \frac{N+1}{2^{N+2}\pi}\left[ 1+\sin\theta_M \sin \theta \cos (\phi-\phi_M)+\cos \theta_M \cos \theta  \right]^N, \\
|\langle \tilde \theta,\phi_M|\theta ,\phi \rangle|^2 &=& \frac{1}{2^N}
\left[ 1+\sin\tilde \theta \sin \theta \cos (\phi-\phi_M)+\cos \tilde \theta \cos \theta  \right]^N,
\end{eqnarray}
we can express Eq. (\ref{EMeanFidNprior}) as
\begin{eqnarray}
\bar F (\theta_M,\phi_M) &=& {\rm const.}\int d\Omega  \left[ 1+\sin\theta_M \sin \theta \cos (\phi-\phi_M)+\cos \theta_M \cos \theta  \right]^N \nonumber \\
&& \times
\left[ 1+\sin\tilde \theta \sin \theta \cos (\phi-\phi_M)+\cos \tilde \theta \cos \theta  \right]^N e^{\kappa \cos \theta }
\nonumber \\
&=& {\rm const.}\int d\Omega \sum_{k=0}^{N}\sum_{k'=0}^{N} {N \choose k}{N\choose k'} (1+\cos \tilde \theta \cos \theta)^k (1+\cos \theta_M \cos \theta)^{k'} \nonumber
\\ && \times
\sin^{N-k}\tilde \theta \sin^{N-k'}\theta_M \sin^{2N-k-k'}\theta \cos^{2N-k-k'}(\phi-\phi_M) e^{\kappa \cos \theta } .
\label{EMeanFidNpriorP}
\end{eqnarray}
We can integrate over $\phi$ using
\begin{eqnarray}
\int_{0}^{2\pi} \cos^k \phi \ d\phi &=& 0, \qquad k \quad {\rm odd}, \\
\int_{0}^{2\pi} \cos^{2n} \phi \ d\phi &=& \frac{2\pi}{2^{2n}}{2n \choose n},
\label{EIntc2n}
\end{eqnarray}
so that after rearranging the summations and expanding the powers of the brackets we find
\begin{eqnarray}
\bar F (\theta_M,\phi_M) &=& {\rm const.} \sum_{s=0}^N 2^{2s}{2N-2s \choose N-s} \sum_{-{\rm min}(s,N-s)}^{{\rm min}(s,N-s)} {N \choose s+r} {N\choose s-r}
\sin^{N-s-r}\tilde \theta \sin^{N-s+r}\theta_M \nonumber \\
&& \times \sum_{l=0}^{s+r}{s+r \choose l}\cos^l \tilde \theta \sum_{l'=0}^{s-r}{s-r \choose l'}\cos^{l'}\theta_M \int_0^\pi \cos^{l+l'}\theta \sin^{2N-2s+1}\theta e^{\kappa \cos \theta} d\theta .
\label{EMeanFidNprior-I}
\end{eqnarray}
The last integral can be expressed by a substitution of $\cos \theta$, leading to
\begin{eqnarray}
\bar F (\theta_M,\phi_M) &=& {\rm const.} \sum_{s=0}^N 2^{2s}{2N-2s \choose N-s} \sum_{-{\rm min}(s,N-s)}^{{\rm min}(s,N-s)} {N \choose s+r} {N\choose s-r}
\sin^{N-s-r}\tilde \theta \sin^{N-s+r}\theta_M \nonumber \\
&& \times \sum_{l=0}^{s+r}{s+r \choose l}\cos^l \tilde \theta \sum_{l'=0}^{s-r}{s-r \choose l'}\cos^{l'}\theta_M
\sum_{j=0}^{N-s}(-1)^j {N-s \choose j}(l+l'+2j)!
\nonumber \\
&& \times \sum_{p=0}^{l+l'+2j}\frac{(-1)^p}{\kappa^{p+1}(l+l'+2j-p)!}\left[ e^\kappa -(-1)^{l+l'+2j-p}e^{-\kappa} \right].
\label{EqTildeThetaM2}
\end{eqnarray}
We find the optimum values $\tilde \theta(\theta_M,\kappa)$ numerically by finding $\tilde \theta$ that maximizes Eq.~(\ref{EMeanFidNprior-I}), or equivalently Eq.~(\ref{EqTildeThetaM2}). These optimum values are presented for various numbers of particles in Fig.~\ref{FigFThetaN}. Unlike for single qubit teleportation, there are multiple values of $\langle n \rangle$ which give a best estimate $\tilde{\theta}$ for a measured value $\theta_M = \pi$ which are not identically 0 or $\pi$.
\begin{figure}[t]
\includegraphics[width=0.8\columnwidth]{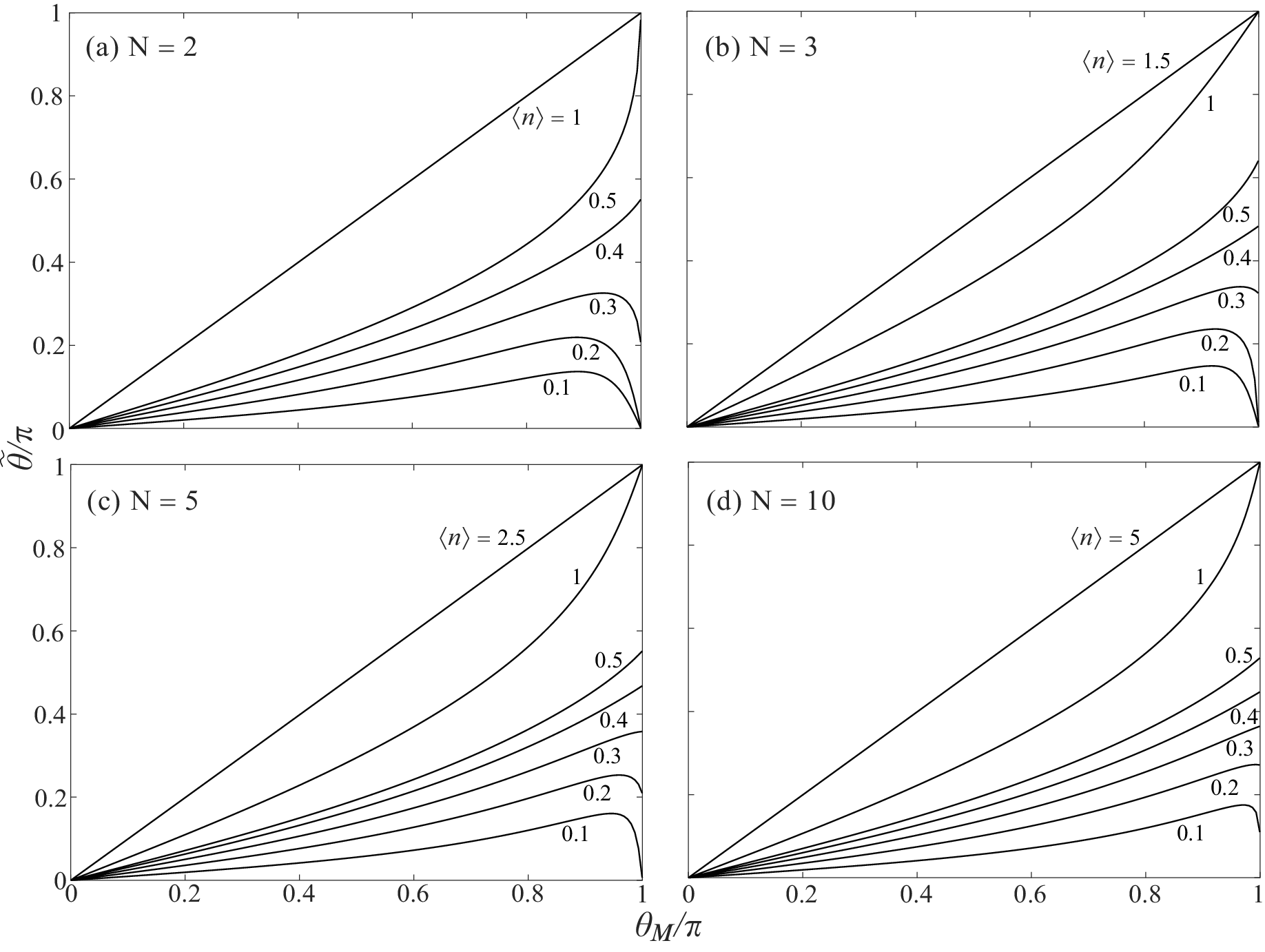}
\caption{Estimated angle $\tilde \theta$ as a function of the detected angle $\theta_M$ for different numbers of qubits $N=2,3,5$ and $10$, and various prior distributions with different mean numbers of excitations $\langle n \rangle$.}
\label{FigFThetaN}
\end{figure}
Averaging $\bar F(\theta_M,\phi_M)$ over all values $\theta_M$ and $\phi_M$ can be expressed as
\begin{eqnarray}
\langle F \rangle &=& \int d\Omega_M \bar F(\theta_M, \phi_M) P(\theta_M,\phi_M) \nonumber \\
&=&  \int d\Omega \int d\Omega_M |\langle \tilde \theta,\phi_M|\theta ,\phi \rangle|^2 P(\theta_M,\phi_M|\theta, \phi ) P(\theta , \phi).
\end{eqnarray}
Using similar steps as before, we simplify three of the integrals to obtain
\begin{eqnarray}
\langle F \rangle  &=& \frac{(N+1)\kappa}{2^{4N+2}\sinh \kappa} \sum_{s=0}^N 2^{2s}{2N-2s \choose N-s} \sum_{-{\rm min}(s,N-s)}^{{\rm min}(s,N-s)}
 {N \choose s+r} {N\choose s-r}
\sum_{l=0}^{s+r} {s+r \choose l} \sum_{l'=0}^{s-r} {s-r \choose l'} \nonumber \\
&& \times \sum_{j=0}^{N-s}(-1)^j {N-s \choose j}(l+l'+2j)!
\sum_{p=0}^{l+l'+2j}\frac{(-1)^p}{\kappa^{p+1}(l+l'+2j-p)!}\left[ e^\kappa -(-1)^{l+l'+2j-p}e^{-\kappa} \right] \nonumber \\
&& \times \int_0^{\pi} \cos^l \tilde \theta (\theta_M,\kappa) \cos^{l'}\theta_M \sin^{N-s-r}\tilde \theta(\theta_M,\kappa) \sin^{N-s+r+1}\theta_M d\theta_M .
\label{EqBarFTildeThetaM}
\end{eqnarray}

\end{widetext}


%

\end{document}